# Overview of Visualization Tools for Web Browser History Data

Abdelhakim Herrouz[1], Chabane Khentout[2] and Mahieddine Djoudi[3]

[1]Department of Computer Science,
University Kasdi Merbah of Ouargla, Algeria

[2]Laboratoire des Réseaux et des Systèmes Distribués,
University Ferhat Abbas of Sétif, Algeria

[3]Department XLIM-SIC UMR CNRS 7252 & TechNE Research Group,
University of Poitiers
Téléport 2, Boulevard Marie et Pierre Curie, BP 30179
86960 Futuroscope Cedex, France

**Abstract**

Nowadays, the Web has become one of the most widespread platforms for information change and retrieval. As it becomes easier to publish documents, as the number of users, and thus publishers, increases and as the number of documents grows, searching for information is turning into a cumbersome and time-consuming operation. Because of the loose interconnection between documents, people have difficulty remembering where they have been and returning to previously visited pages. Navigation through the web faces problems of locating oneself with respect to space and time. The idea of graphical assistance navigation is to help users to find their paths in hyperspace by adapting the style of link presentation to the goals, knowledge and other characteristics of an individual user. We first introduce the concepts related to web navigation; we then present an overview of different graphical navigation tools and techniques. We conclude by presenting a comparative table of these tools based on some pertinent criteria.

## 1. Introduction

Due to the rapid growth of the Web, sites appear and disappear, content is modified and it becomes impossible to master their organization. In fact, the navigation process on the web is confronted by three major problems. On one hand, the nature of the environment itself imposes some disadvantages: Internet is a network of worldwide level, constantly changing and non-structured. Next, users generally have difficulties in constructing a mental navigation outline. At last, the computer-aided tools for navigation offered by different classical software do not satisfy the user needs and sometimes contribute, paradoxically, to make the navigation process more confusing [1]. The conclusion from the analysis of these problems is to develop new computer aided tools for navigation. These tools will have to be able to address the following two main questions usually asked by the user: "which link to follow?" and "how to retrieve this page?". The Internet representation tools and the user navigation path visualization are certainly answers provided by the current and feature developments [2].

The remaining sections of the paper are organized as follows. First we discuss the general problems related to the navigation on the Web and other difficulties encountered. After that, we present and compare the main computer aided tools for navigation available in the literature.

## 2. Browsing Model

Most of the tools are designed to improve navigation through the information space and enable people to find what they are looking for more easily. It is important to distinguish between browsing and searching for information in a large information space like the Web. They are very different activities which require different support tools. Browsing is largely an explorative activity, usually with no planning or specific goals, with useful results dependent on serendipity. At present, the Web supports two major forms of browsing: link-following and directories. Browsing by link-following uses the fundamental Web function of hyperlinks connecting pages that can be explored using the standard browser application. However, browsing hyperlinks between pages can often be frustrating and unproductive, as it is all too easy to get lost in the complex topologies of links as there is a lack of navigational cues indicating where you started from, where you are at present or where you can go onto. Users waste much time wandering through Web sites without finding anything of interest or gaining any useful insight. After a while wandering lost through the Web, users are often forced to go back to the entrance point and start again. Generally, we distinguish three navigation models: spatial, semantic and social [3].

Spatial navigation is based on the analogy with the real world and in particular our knowledge of the space (proximity notion, alignment, etc.). It is especially used in virtual reality systems but also in information systems. This navigation model sets goals to be reached - for example find certain information - and from the user view of point, it raises two questions [4]: (1) Where am I? (2) Where is …? How do I go to …? Does … exist somewhere in the space?

In addition to the component temporal (past, present, future) that plays a basic role in navigation. The above mentioned questions identify the spatial navigation of the activity - paths and places -. Nevertheless, this spatial aspect underlines another important parameter: the traversal means.

Semantic navigation describes the user behaviors when he/she moves in the information space according to the information attributes that are presented (similitude, value, etc.). Its implementation is fundamental because it allows the navigating user to accomplish practically, all the required tasks. This navigation model is used with hypertext systems (paths through the hyperlinks) but does not exploit the characteristics spatial of information. It is used by the users browsing the Web. In fact, the movement from one document to another is done by a click of the mouse on an object and the location of the latter has no effect on the destination of the link [4].

The third model is social navigation that is based on the exploitation of information about other users. This type of navigation supposes that the users share the same information space [5], [6].

At the end we note that these three models do not exclude each other and the combination of several navigation types allows the user to benefit from a better interaction with his/her information space [4].

## 3. Web Browsing Difficulties

The Web combines difficulties that are usually present whenever a huge information system is used, with conceptual difficulties linked to the choices and the progression through heterogeneous information. The difficulties encountered during navigation are various but they can be classified into two general types: the disorientation and the cognitive overhead [7].

**Disorientation:** Disorientation [8] can be defined as the mental state of feeling lost when navigating in hypertext systems. It is a psychological state resulting from problems in constructing pathways across a hypertext. The indications of disorientation based on the self-reported research data show that users:
(1) do not know where to go next;
(2) know where to go but not how to get there;
(3) and, do not know where they are in relation to the overall structure of the document.

Consequently, they may become frustrated, lose interest, and experience a measurable decline in efficiency.

**Cognitive overhead**: The cognitive overhead happens with a user who has only a screen to work with. This user has to know the information shown is associated with what. Many decisions have to be taken while going through a hypermedia: which link to follow, how to retrieve the ones that are of interest among the links already visited or to be visited.

The user should be able to find the information being searched while moving from one page to another by following the different links. These tasks of searching for what is needed require accessing the information in smart way. This means that we need to have the capabilities to go from one place to another, identify the document reached, evaluate it, to save it or memorize its address, and related to other documents and information.

It is very common to notice that during the use of hypermedia, the user, after few minutes of search, does not know where he really is with respect to the different notions he went through. We reach a point where we start to move from one page to another or from one site to another without gaining anything new even if some of pages and/or site may contain relevant information. This is not going to improve the knowledge of the learner [9].

Working with the Web may lead the user, from one link to another, to a page that has very little to do with the subject being searched for. The information read that is not related to a specific cognitive project is forgotten very quickly. Meanwhile, we forget other pages that we have consulted earlier which contained information that is of interest to us. We activated a link that we taught it would allow us to get more information about the topic. This action took us further away from the subject because we kept following other links. Before we noticed it, we lost track the pages that interest us. After a half-hour of search, we turn off our computer with the impression that we went through a lot of material without learning anything new.

## 4. Browsing Help

Navigation help takes essentially two forms. The first way is concerned with the construction of web sites. A construction method should be adapted to make it easy for the user to access and search the sites. For example, in Quarteroni et al. [9], the author proposes to limit the depth decomposition of the page to four levels. This means, only three nodes can be active at the same time. In addition, each screen should have about five active links. In order to be clear and efficient, links to general ideas of dependant information are favored. This approach of construction will result into hypermedia with a simple and efficient structure. The inconvenience of this method is that the user has to split for example a design of a complete course into subsections that are accessed separately. But we can always link these subsections to each other indirectly.

The second way is to provide a set of computer-aided tools that will allow the client to navigate the web with ease using his/her preferred browser. The general browsers, Firefox or Internet Explorer propose some functionality such as history, and bookmarks but these kinds of help are insufficient for user needs. In addition, the users of a hypertext system create different representations. Many computer-aided systems that help the users while browsing the Internet have been proposed in the literature [10], [11], [12], [13], [14], [15], [5], [6]. After we present the principal computer-aided navigation tools, we give a comparative table based on some essential criteria of usage and functionality.

## 5. Visual Map for Browsing

The development of a graphical map and its use as a computer-aided tool for web browsing is based on the studies of cognitive processes that happen during the navigation of distributed hypermedia. It is a graphical representation at the same time of conceptual and geographical search path followed by a user while searching for a particular topic. The Navigation map that we designed is based on the idea used in conceptual maps [16].

A conceptual map is a new way of representing the relationship between a set of knowledge and the nature of this relationship. It is a graphical representation of links among different concepts about the same topic. It should evolve with the knowledge of the trainee.

The conceptual map is also a computer-aided tool for navigation. It allows a hypertext reader to see on the screen the titles of information units and the links that connect them in a form of a network. It is drawn with a goal in mind, within well-defined references, and according to a graphical representation suitable for browsing problem.

## 6. Classification of Visual Representations

Browsing the Web implies the manipulation of huge amount of information. The major role of the graphical interface of system developed for this purpose is to make this information easy to comprehend by the users. This is based mainly on the graphical representation of the different pieces of information and the relations connecting these pieces together. The graphical interface between the users and the system is a way to construct the image of the system. A review of the literature indicates the existence of many graphical representations. So, it is necessary to study and classify these different representations.

The taxonomy developed by Tweedie [17] is based on the notion of the user's actions. The classification proposed emphasizes the nature of actions (direct or indirect selections), their levels (single, group, and attributes and objects integrity) and their effect on the graph, on the representation and the transformation or organization of the objects selected.

The study proposed in [18] classifies representation techniques in five categories: geometric, network based, hierarchy, pixel oriented, and iconic. This approach has the disadvantage of mixing construction and graphical tools used as a classification criteria, which makes it very difficult to characterize some systems.

The approach described in [15] is based on the type of data represented and the low level task performed by the user on this data. The author then listed different graphical representations used for each type of data. He also identifies seven task types that the graphical representation should favor. The high level tasks that are independent of the data being manipulated are: general view of the information, zooming, filtering, getting the details, link representation together, having a history of actions performed, and extracting part of the information so that it can be used by other applications. Three of these points (general view of the information, zooming, and getting the details) are considered during the conception of the representation.

In [19], the authors propose to characterize the graphical representation based on a chosen point of view about the data but not on the type of data. A point of view is defined by deciding what is necessary out of the data that should be given to the users based on his needs to perform his task in a satisfactory manner. If we are unable to characterize in a precise way the object's activities, then the graphical representation should be flexible enough to detect one or many points of view that are suitable to accomplish the task. For a set of data, we might have more than one point of view depending on how the data is considered. These points of views might complement each other for the purpose of the user's activities. So it is necessary to be able to represent simultaneously many views which means we should choose a graphical representation guided by multiple points of view. This corresponds to multiple views discussed in [20] and [4]. This multiplicity should be taken as a factor during the design of an interface that can adapt itself to different tasks [1], [21].

## 7. Web Browser History Data Overview

**NaVir:** In order to allow the user to keep track of time and to know where he/she is, we have designed and implemented a computer-aided system for virtual navigation of the web called NaVir. This system which is implemented in Java can be used with any browser (Firefox, Internet Explorer or other). The main screen is made up of many windows. Its kernel is made up of two important modules: one is to collect the different URL addresses and the other is to build and interact with the graphical map and the management of navigation time.

In order to guarantee that our system is independent of the browser, the way we recuperate the addresses of the sites/pages visited is using a proxy server. This proxy server seats in between web clients and information servers using different protocols. It is used to pass the information from one end to the other. Each user's request is sent by the client to the proxy server which will respond directly if it has the information in its cache, or it will pass the request to the destination server. The proxy server keeps a copy of each document it sends in its cache. This copy is kept for variable amount of time. This way, if a document is requested and is available in the cache of the proxy, there is no need to get it from a distant server. The memory cache management is done based on the following parameters: date of the last time when the document was updated, maximum time that a document can spend in the cache and for how long has the document been in the cache without being used. This service, which is transparent to the user, offers the responses to the user requests in an efficient way. It also reduces the traffic on the network. Navigation time by the users is included. It is an excellent tool to model the user behavior during navigation. NaVir is being used to facilitate the learning process within a platform for distance education on the Web [20].

**Nestor:** NESTOR [22] was developed at CNRS-GATE laboratory. It is a Web browser that draws interactive web-maps of the visited Web space during navigation: the objects that show on Nestor maps are the visited web documents and the links that have been used to reach them. The web-maps are hybrid in the sense that users can add objects of their own – concepts, links, personal documents, organizers – and progressively evolve the maps into concept-maps. The maps are interactive in the sense that they provide direct navigation back to the represented objects, and allow for a full set of drag-and-drop operations aimed at structuring the information extracted from the Web: Nestor combines graphical Web navigation and mind-mapping features. Nestor is also collaborative software that enables small groups of people to share their navigation experience. We could say that Nestor promotes a constructionist approach to Web information mapping. Nestor is a complete and excellent navigator. It is a very good tool to build the navigation map. However, the client software is platform dependent; it runs only on top of Microsoft Internet Explorer on Microsoft Windows platforms.

**Broadway:** The navigation helper Broadway (a BROwsing ADviser reusing path WAYs) is a server that keeps track of document requests made by the customers by saving them. Broadway can be accessed by a group of users and supports indirect cooperation. It uses a reasoning system based on cases to advise a group of users on the interesting pages to visit according to the path that the group has already traversed. It establishes the reasoning system from cases that confirm to a flexible and generic framework formed by an index model of different situations. It helps a user who is navigating on the Web and facilitates the task of searching information on this hypermedia. The interaction of the user with Broadway is assured by the assistance of two means: the tool bars and the controller. Broadway has an open and well-adapted architecture to the Web [18].

**Footprints:** This tool presents a visualization technique modeled by a graph where every node symbolizes a page. The nodes are linked together by links representing paths traversed by the users. In addition, different colors are assigned links to show their usage frequency. The user can therefore visualize the graph to locate himself and choose a link to follow by a simple click on the graph. Footprints is based on the principle that if several users followed a particular link, then this link is interesting to recommend. The system displays the more frequently visited set of pages from the current page. Besides, Footprints uses the HTTP logs of a specific server to construct the graph of users' searched paths [15].

**Hypercase:** The technique used in Hypercase [12] is the only known example of map adaptation. This technique supports local and global orientation by adapting the form of local and global maps to the didactic or information goal of the users. Hypercase represents and uses knowledge about possible goals for goal adaptation. Hypercase uses a case-based approach and a neural network technology to store in the database of cases several typical navigation paths for each of the didactic goals. Using this knowledge, the system can find the most similar standard path (and thus the most probable didactic goal) for the navigation path of a real student supplied as an input to the case-based mechanism. When the student requests help, Hypercase can show where he/she is located in the hyperspace by drawing a wide-area or local area hierarchical map. As the root of the hierarchy the system uses the "central node" of the hyperspace (which is computed by a special method) for the wide-area map and the closest node of a deduced standard path for the local-area map.

**Letizia:** Letizia [23] is a behavior-based interface agent which doesn't require the user to provide an explicit initial goal. Rather, it attempts to infer the goal from the user's actions. It tracks user behavior and attempts to anticipate items of interest by doing concurrent, autonomous exploration of links from the user's current position. Letizia simply suggests a list of hyperlinks ordered by preference, and can give the user a reason for the recommendation upon request. Letizia doesn't require the user to evaluate the previous searches as successful or unsuccessful, but instead applies heuristics, learning the user's interest through the user's behavior. The subjects are stored as lists of keywords. Using this representation of user interest, it performs a best first search, following links and evaluating against the subjects of interest, eliminating dead end links. The user's previous interests are stored and persist while the user browses over time, and they decay by a factor over time. Let's Browse [24] is the multi-users version of Letizia. It allows group navigation.

**WebView:** WebView is an add-on window to Netscape Navigator that presents an automatically generated graphical overview of the user's browsing paths. It provides a variety of facilities for navigational shortcuts, and it allows the user to tailor the display of a large set of pages. As with conventional systems, clicking on the text-title alongside any page makes Netscape navigate to the page. It also detects the title and URL of the page, and these are (optionally) displayed alongside the thumbnail. Because some thumbnails may be difficult to distinguish from others (such as a site's pages that follow a standard look), it provides larger views: mousing over any miniaturised thumbnail causes it to zoom to approximately four times the size [25].

**PadPrints:** PadPrints is a browser companion called PadPrints that dynamically builds a graphical history-map of visited web pages. PadPrints relies on Pad++, a Zooming User Interface (ZUI) development substrate, to display the history-map using minimal screen space. PadPrints functions in conjunction with a traditional web browser but without requiring any browser modifications. Also in PadPrints a node in the hierarchy displays the title of the web page and a small picture associated with the page. Finally, the systems construct the hierarchy as users traverse links from one page to another, as opposed to prebuilding a hierarchy for a single website. The PadPrints browser companion monitors and controls the web browser. When users access pages from the web browser those pages are added to the PadPrints display. Pages are added as children of the current node in the hierarchy, unless that page is already present in the hierarchy. A single click on a page in the PadPrints display sends the browser to the corresponding URL [26].

**WebWatcher:** WebWatcher [27] uses the current page and a set of key words provided by the user at the start of the search. Then, it highlights the recommended hyperlinks of the current page. It is implemented according to a similar architecture of an HTTP server proxy. It examines and modifies the links of the visited pages so that it redirected them to the same server. This way, WebWatcher can therefore follow the users during their navigation. WebWatcher requests an initial goal from the user, and the e-mail address to keep track of the user's interests. WebWatcher enhances the basic Web browser page with: a menu bar above the page to communicate with the agent, a list of new hyperlinks found to contain the words in the goal, hyperlink recommendations and highlighted hyperlinks. The original prototype was implemented for Mosaic users. The actual learning of the system was acquired by logging a user's successful and unsuccessful searches as training data. It suggests an appropriate hyperlink based on the current web page viewed by the user and the user's information goal.

**WBI:** WBI [28] is another single-user computer aided tool that saves the navigation of a user and then analysis it to extract typical sequences that are produced often. This allows the optimization of the user navigation. WBI proposes the final page of a sequence as soon as the user displays the first page. It is based on the technique of proxy server and has a modular architecture allowing the collaboration of different agents. WBI provide to collect the navigation data of a user in the Web, capturing the entire exchange of information between these two means of usability evaluation, without access restrictions to the information. Moreover, it contains a low transparency to the user. Still, these tools present a few problems: (a) all information necessarily passes through an intermediary, slowing navigation; otherwise, depending on the quality of the connection, this can become a problem; (b) all information required for evaluation is captured with the user's personal information; but to guarantee the data's security (even if not kept or used) generates doubts; (c) the information ends up becoming homogenous due to the lack of contextualization of the actions, as there is no distinction between the type, form, or use of each action.

**Yan et al:** The system design facilitates the analysis of past user access patterns to discover common user access behavior. The information can then be used to improve the static hypertext structure, or to dynamically insert links to web pages. In the offline module, the preprocessor periodically extracts information from user access logs to generate records of users sessions. One record is generated for each session in the logs. The record registers the access patterns exhibited by the user in that session. Records are then clustered into categories, with "similar" sessions put into the same category. The online module performs dynamic link generation. When a user requests a new page, the module tries to classify his current partial session record against one or more of the categories obtained offline. The top matching categories are identified, and links to unexplored pages contained in these categories are inserted at the top of the page shipped back to the user. Experimental results obtained by analyzing real user access logs show that indeed clusters of user access patterns exist. Further, some of these clusters are not apparent from the physical linkage of the pages, and thus would not be identified without looking at the logs [29].

## 8. Comparative Study

Comparison Criteria: The different visualization tools for web browser history data are difficult to compare because of the variety of goals and contexts. In the framework of our applications constraints, we compare the existing tools based on the following six points [1]:

- Visualization technique used: It depends on how advanced is the offered visualization technique (map, tree, etc.).
- Annotation: The system proposes the possibility to annotate the links.
- Interaction: The capacity of the system to react to different interactions of the user.
- General assistance: The system allows multi-sites or a specific hypermedia.
- Open: The tool can change and evolve according to different strategies;
- Independent: The independence from the navigators.

**Comparative Table:** The following table summarizes the characteristics of these visualization tools. In the columns, we use the following symbols:
- - : for No
- √: for Yes
- Z: map visualization (Zoom)
- A: possibility to Annotate links or content
- H: degree of Help
- O: degree of Opening
- T.m: Time management
- I: Independency of tool to the web browser

Table 1: Comparative Table of Browser History Data

| *Tool* | *Z* | *A* | *H* | *O* | *T.m* | *I* |
|---|---|---|---|---|---|---|
| *NaVir* | √ | √ | √ | - | √ | √ |
| *Nestor* | √ | √ | √ | √ | - | - |
| *Broadway* | - | - | √ | √ | - | √ |
| *Footprints* | √ | √ | - | - | - | ? |
| *Hypercase* | - | - | - | - | - | ? |
| *Letizia* | - | - | √ | √ | - | ? |
| *WebView* | √ | - | ? | - | - | - |
| *PadPrints* | √ | - | - | ? | - | √ |
| *WebWatcher* | - | - | √ | √ | - | √ |
| *WBI* | - | √ | √ | √ | - | √ |
| *Yan et al.* | - | - | - | - | - | √ |

As can be seen from the table above, we can notice the following:
- Four out these tools offer the annotation possibility: NaVir, Footprints, Nestor and WBI.
- The systems NaVir, Broadway, Letizia, Nestor, WebWatcher and WBI allow multi-sites assistance. They aim therefore for assistance on the user side by using the proxy server technique or the links redirection. On the other hand, Footprints, Hypercase and Yan's approach aim to a restricted assistance to a specific server. They are therefore linked to a special hypermedia.
- NaVir is different from the other tools because it gives the user the possibility of managing the navigation time spent and knowing how much time is spent on each page or a site.
- Nestor Web browser uses a specific navigator (Microsoft Internet Explorer); so its use is limited to a precise platform (Microsoft Windows).

## 9. Conclusion

In this paper, we presented a non-exhaustive list of the available visualization tools for web browser history data. Through this study, we established some objective criteria for comparison. Based on these criteria, we gave a comparative table of these different tools. We are currently developing client software to build a navigation map. The system is based on multi-agent technology and it draws interactive Web maps while we are surfing the Web [30].